# Boundary Conditions at the Walls with Thermionic Electron Emission in Two Temperature Modeling of "Thermal" Plasmas

L. Pekker and N. Hussary

Victor Technologies, West Lebanon, NH 03781, USA

## Abstract

In this paper we propose new boundary conditions at the hot walls with thermionic electron emission for two-temperature thermal arc models. In the derived boundary conditions the walls are assumed to be made from refractory metals and that the erosion of the wall is small and, therefore, is not taken into account in the model. In these boundary conditions the plasma sheath formed at the electrode is considered as the interface between the plasma and the wall. The derived boundary conditions allow the calculation of the heat flux to the walls from the plasma and consequently the thermionic electron current that makes the two temperature thermal model self consistent.

## 1. Introduction

The formation of the plasma sheath at cathodes and anodes plays a fundamental role in the structure of the cathode spot and anode attachment, electrode erosion process, thermionic emission, heat flux to the wall from the plasma, and other electrode processes. However, one fundamental question remains open: what boundary conditions should be used at the electrodes for hydrodynamic modeling of thermal plasmas which would take into account the plasma sheath formed at the wall? In previous studies of high-pressure plasmas (the plasma pressure is as large as or larger than atmospheric pressure) different sets of boundary conditions at the cathode with thermionic electron emission were constructed disregarding the cathode sheath, see for example local thermodynamic equilibrium models $(T_e = T_h)$ [1 - 5] and two temperature $(T_e \neq T_h)$ models [6 - 18] and references therein; $T_e$ is the electron temperature and $T_h$ is the temperature of heavy particles, ions and neutrals. In recent work [19] the authors constructed sets of boundary conditions at the floating and biased walls for the case of cold walls with no thermionic electron emission or no erosion of the wall. In this work they considered the plasma in a two temperature

-----------

Corresponding authors is Leonid Pekker, Tel. +1 (661) 378-4811 (c) Leonid.Pekker@gmail.com

hydrodynamic approximation, $T_e \neq T_h$, and the plasma sheath formed at the wall as the interface between the plasma and the wall, and used Godyak's collisional sheath model [20, 21]. In [19] it was demonstrated that using these boundary conditions in modeling of high-pressure arcs may lead to much larger heat fluxes to the wall and to significantly cooler arcs compared to models that ignore the sheath at the wall.

In the present paper we extend model [19] to the case of hot biased walls (electrodes) with thermionic electron emission where the walls are assumed to be made of refractory metals, and that erosion of the wall is small and can be neglected in the model. In this model we expand the Godyak's sheath model [20, 21] to the case of thermionic electron emission. We construct the boundary conditions at the electrode (cathode or anode) with thermionic electron emission for electric potential in Section II and for the electron and heavy particle energy equations in Section III. An algorithm of implementation of the suggested boundary conditions is presented in Section IV. To illustrate the effect of these boundary conditions, we consider the case of the virtual cathode in argon plasma, Section V. In Section VI we propose a model of a cathode arc in the case where the heat flux from the plasma to the wall is balanced by the energy flux that the thermionic electrons bring back to the plasma. The conclusions are given in Section VII.

## II. Boundary conditions at the wall for the electric potential

One of the important issues in hydrodynamic modeling of thermal plasmas is what boundary conditions for electrical potential should be used at the wall which would take into account the plasma sheath formed at the walls. This issue is considered in this section.

Since the plasma is assumed to be quasineutral, the Poisson equation in the plasma can be written as

$$\nabla(\sigma \cdot \nabla\varphi) = 0. \qquad (1)$$

The boundary conditions for this equation have to be written not at the wall, where the quasineutrality condition, $|n_e - n_i|/n_i \ll 1$, is not valid, but rather at the plasma-sheath interfaces where the plasma is quasineutral, Figs. 1 and 2.

In assumptions of two-temperature "thermal plasma", $T_e \neq T_h$, and no slip temperature at the wall, the temperature of heavy particles, ions and neutrals, at the wall can be taken as $T_{sur}$ [1 - 18]; $T_{sur}$ is the wall surface temperature. The reason for this is because the ions that reach the wall are thermalized to the temperature of the wall (same holds true for neutrals). The ions recombine with electrons at the wall and make their way back to the plasma as neutrals where they are immediately ionized by electrons. The case of the electrons is different. The plasma sheath formed at the wall, Fig. 1, rebounds most of the "plasma" electrons back to the plasma; basically separating the "plasma" electrons from the wall. At the same time the thermionic electrons emitted by the hot cathode (anode) are accelerated in the sheath to energies much larger than the plasma electron temperature. They enter the plasma and heat the "plasma" electrons. Since the thermalization rate of electrons with themselves is much faster than the electron - heavy particle energy exchange rate, the temperature of electrons at the cathode has to be larger than $T_{sur}$.

As in [19] we will (1) consider the plasma sheath as the interface between the wall and the plasma, (2) assume that the plasma at the wall is singly ionized, (3) assume the potential in the sheath is monotonically decreases from the plasma side to the wall, (4) set $T_e \gg T_h = T_{sur}$. (5) We will also assume that the thermionic electrons pass through the sheath collisionlessly transferring their momentum and energy far from the sheath in the plasma; in other words the thickness of the cathode sheath is much smaller than the transport mean free path for thermionic electrons,

$$L_{cath}^{sheath} < \lambda_{e-therm-mfp}. \qquad (2)$$

*2A. Cathode sheath*

First consider the case where the electrode is a cathode with thermionic electron emission. To obtain the sheath potential at the cathode, Fig. 1, the total current density at the cathode surface can be written as:

$$j = j_i - j_{e-plasma} + j_{e-therm} \qquad (3)$$

where the ion current density in the sheath is:

$$j_i = e n_p V_s,  \quad (4)$$

$n_p$ is the plasma density at the plasma-sheath interface,

$$V_s = V_{i-sound} \left(1 + \frac{\pi r_{De}}{2\lambda_{i-mfp}}\right)^{-1/2} \quad (5)$$

is the ion velocity at which the ions enter the sheath [20, 21], $V_{i-sound} = \sqrt{k_B T_e / M}$ is the ion sound speed, $M$ is a mass of a heavy particle, $\lambda_{i-mfp} = 1/(n_n \cdot \sigma_{i,n})$ is the ion transport mean free path, $\sigma_{i,n}$ is the charge-exchange cross section, the dominated ion - neutral momentum transfer process in the sheath [20, 21], $r_{De} = \sqrt{\varepsilon_0 k_B T_e / n_p e^2}$ is the electron Debye radius; the electron plasma current density can be expressed as:

$$j_{e-plasma} = e n_p \exp\left(-\frac{e\varphi_{cath}}{k_B T_e}\right) \sqrt{\frac{k_B T_e}{2\pi m_e}}, \quad (6)$$

$\varphi_{cath}$ is the sheath potential drop between the plasma and the cathode; and the current density of "thermionic" electrons is

$$j_{e-therm} = A T_{sur}^2 \exp\left(-\frac{e(\varphi_{work-func} - \Delta\varphi_{Schot})}{k_B T_{sur}}\right). \quad (7)$$

In Eq. (7) we use the Richardson law with the Schottky correction factor

$$\Delta\varphi_{Schot} = \left(-\frac{eE_{sur}}{4\pi\varepsilon_0}\right)^{1/2} \quad (8)$$

describing the decrease of the effective work function of materials in strong electric fields; $E_{sur}$ is the electric field at the surface of the wall, and $A$ depends on the cathode material.

Substituting Eqs. (4 - 6) into Eq. (3) and after solving for $\varphi_{cath}$ we obtain

$$\varphi_{cath} = -\frac{k_B T_e}{e} \ln\left[\sqrt{\frac{2\pi m_e}{M\left(1+\frac{\pi r_{De}}{2\lambda_{i-mfp}}\right)}}\left(1 - \frac{j-j_{e-therm}}{j_i}\right)\right] \qquad (9)$$

where

$$j_i = en_p V_p = en_p \sqrt{\frac{k_B T_e}{M\left(1+\frac{\pi r_{De}}{2\lambda_{i-mfp}}\right)}}; \qquad (10)$$

the thickness of the cathode sheath is on the order of $r_{De}$. Because the plasma electron number density in the sheath is assumed to be small, on the order of $n_p exp(-e\varphi_{cath}/k_B T_e)$, and $e\varphi_{cath} \gg k_B T_e$, the friction of ions with plasma electrons can be neglected relative to the ion-neutral friction, as well as the friction of the ions with the thermionic electrons, see Section III. The case when $r_{De} \ll \lambda_{i-mfp}$ corresponds to the collisionless Bohm's sheath [22], where the ions are freely accelerated in the sheath, and the case $r_{De} > \lambda_{i-mfp}$ corresponds to the collisional sheath [20, 21], where the ions move in the sheath in the charge exchange regime and the $\lambda_{i-mfp}$ is independent of the ion velocity [21].

It is worth noting that in the case of very high gas pressure, where $\lambda_{i-mfp} \gg r_{De}$, $V_s < \sqrt{k_B T_h/M}$, and the ions move in the sheath in the mobility (not charge exchange) regime ($\lambda_{i-mfp}$ in the mobility regime is dependent on the ion velocity), and Eq. (9) should be modified accordingly [21]. This case is not considered in the present paper.

To obtain $j_{e-therm}$ at a given plasma and wall parameters: $j, n_p, n_n, T_e$ and $T_h = T_{sur}$ we have to calculate $E_{sur}$ by solving the Poisson equation in the cathode sheath. The Poisson equation in the plasma sheath formed at the electrode with thermionic electron emission can be written as:

$$\frac{d^2\varphi}{dx^2} = \frac{e}{\varepsilon_0}\left(n_p exp\left(\frac{e\varphi}{k_B T_e}\right) - \frac{n_p}{\sqrt{1-\frac{2e\varphi}{MV_s^2}}} + \frac{j_{e-therm}}{e\sqrt{\frac{2e(\varphi_{cath}+\varphi)}{m_e}}}\right). \qquad (11)$$

In Eq. (11) the first term in the brackets is the density of "plasma" electrons, the second term is the density of ions, and the third is the density of the "thermionic" electrons in the sheath. The $x$-coordinate is directed from the plasma to the sheath, Fig. 1. Eq. (11) can be solved with the following boundary conditions:

$$\varphi(x=0) = 0, \quad \left(\frac{d\varphi}{dx}\right)_{x=0} = -\frac{k_B T_e}{e r_{De}}, \tag{12}$$

where the first condition states that the potential at the sheath from the plasma side is equal to zero, Fig. 1, and the second one is chosen according to the Godyak sheath model [20, 21]. Although Godyak used this boundary condition for the case of no secondary electron emission, it can be also applied for thermionic electrodes. As has been mentioned in [21], the second condition, in fact, describes the "electrostatic wall" separating electrons from the wall. This is reasonable because the density of plasma electrons at the cathode in the model is assumed to be much smaller than in the plasma, $exp(-e\varphi_{cath}/k_B T_e) \ll 1$.

As one can see from Eq. (11) at $x = 0$, the plasma is not quasineutral, and

$$n_e(x=0) - n_i(x=0) = \frac{j_{e-therm}}{e\sqrt{\frac{2e\varphi_{cath}}{m_e}}}. \tag{13}$$

As a result, $\varphi_{cath}$ calculated by the model is smaller than the "real" potential drop between the quasineutral plasma and the wall; this difference $\Delta\varphi$ can be estimated as

$$\Delta\varphi = -\frac{k_B T_e}{e} \ln\left(1 - \frac{j_{e-therm}}{e n_p \sqrt{\frac{2e\varphi_{cath}}{m_e}}}\right). \tag{14}$$

Thus the suggested model is reasonable only if this condition,

$$\frac{\Delta\varphi}{\varphi_{cath}} = -\frac{k_B T_e}{e\varphi_{cath}} \ln\left(1 - \frac{j_{e-therm}}{e n_p \sqrt{\frac{2e\varphi_{cath}}{m_e}}}\right) \ll 1, \tag{15}$$

is well satisfied.

Integrating Eq. (11) using Eq. (12), we obtain the following equation for $\varphi$:

$$\frac{d\varphi}{dx} = -\frac{k_B T_e}{er_{De}} \left( \begin{array}{c} 2\left(\exp\left(\frac{e\varphi}{k_B T_e}\right) - 1\right) + 4\alpha\left(\sqrt{1 - \frac{e\varphi}{k_B T_e}\frac{1}{\alpha}} - 1\right) + \\ 4\beta\left(\sqrt{\frac{e\varphi_{cath}}{k_B T_e} + \frac{e\varphi}{k_B T_e}} - \sqrt{\frac{e\varphi_{cath}}{k_B T_e}}\right) + 1 \end{array} \right)^{1/2}, \quad (16)$$

which can be solved numerically;

$$\alpha = \frac{MV_s^2}{2k_B T_e} = \frac{1}{2\left(1 + \frac{\pi r_{De}}{2\lambda_{i-mfp}}\right)} \quad \text{and} \quad \beta = \frac{j_{e-therm}}{en_p\sqrt{\frac{2k_B T_e}{m_e}}}. \quad (17)$$

Substituting $\varphi = -\varphi_{cath}$ into Eq. (16) we obtain an equation for the electric field at the wall:

$$E_{sur} = \frac{k_B T_e}{er_{De}} \left( 2\exp\left(-\frac{e\varphi_{cath}}{k_B T_e}\right) + 4\alpha\left(\left(1 + \frac{e\varphi_{cath}}{k_B T_e}\frac{1}{\alpha}\right)^{1/2} - 1\right) - 4\beta\sqrt{\frac{e\varphi_{cath}}{k_B T_e}} - 1\right)^{1/2}. \quad (18)$$

Thus, solving the equation set (7) - (9), and (18) at a given $j$, $n_p$, $n_n$, $T_e$ and $T_{sur}$ we can obtain $\varphi_{cath}, j_{e-therm}, \Delta\varphi_{Schot}$, and $E_{sur}$.

As it will be shown in Section V, this set of equations does not always have a solution. At high magnitudes of $j_{e-therm}$ the value in the brackets in Eq. (18) can become negative which leads to imaginary $E_{sur}$. This case corresponds to the formation of a "virtual cathode", where the potential, $\varphi$, in the sheath is not a monotonic function on $x$ as our model assumes, see assumption (3). In this case not all thermionic electrons emitted from the cathode surface reach the plasma, some of them are rebounded back to the cathode that leads to a decrease in the current density of thermionic electrons reaching the plasma. Critical value of thermionic electron current density, $j_{e-therm}^{critical}$, and the critical cathode sheath potential drop $\varphi_{cath}^{critical}$ corresponding to the case of the virtual cathode, can be obtained at a given $j$, $n_p$, $n_n$, $T_e$ by solving the following set of equations numerically:

$$2exp\left(-\frac{e\varphi_{cath}^{critical}}{k_BT_e}\right) + 4\alpha\left(\sqrt{1+\frac{e\varphi_{cath}^{critical}}{k_BT_e}\frac{1}{\alpha}} - 1\right) - \frac{4j_{e-therm}^{critical}}{en_p\sqrt{\frac{2k_BT_e}{m_e}}}\sqrt{\frac{e\varphi_{cath}^{critical}}{k_BT_e}} - 1 = 0, \quad (19)$$

$$\varphi_{cath}^{critical} = -\frac{k_BT_e}{e}\ln\left[\sqrt{\frac{2\pi m_e}{M\left(1+\frac{\pi r_{De}}{2\lambda_{i-mfp}}\right)}}\left(1-\frac{j-j_{e-therm}^{critical}}{j_i}\right)\right] \quad (20)$$

Eq. (19) states that the electrical field at the virtual cathode is equal to zero. Thus, if $j_{e-therm}$ at a given $T_{sur}$ with $\Delta\varphi_{Schot} = 0$, Eqs. (7), is smaller than $j_{e-therm}^{critical}$, then the set of Eqs. (7) - (9), and (18) have a solution and $j_{e-therm}$, $E_{sur}$, $\Delta\varphi_{Schot}$, and $\varphi_{cath}$ can be obtained. In the case where $j_{e-therm}$ with $\Delta\varphi_{Schot} = 0$ is larger than $j_{e-therm}^{critical}$, the case of virtual cathode, we suggest to use $j_{e-therm}^{critical}$ and $\varphi_{cath}^{critical}$ instead of $j_{e-therm}$ and $\varphi_{cath}$.

For the sake of simplicity one may use $k_BT_e/er_{De}$, the electric field at the boundary of the sheath facing the wall, Eq. (12), instead of $E_{sur}$ in the Schottky correction factor, Eq. (8). However, since the electric field at the cathode surface can differ significantly from $k_BT_e/er_{De}$, Section VI, this is not recommend. As it was stressed in Nemchinsky's review paper [23], the experiments on free burning arcs at atmospheric pressure [24] clearly show that $\Delta\varphi_{Schot}$ is fundamentally dependant on the cathode spot current. This will be demonstrated for the cathode spot model in Section VI.

In the case of cold cathode where the thermionic electron emission and erosion of wall are negligibly small the cathode sheath potential drop can be obtained by dropping $j_{e-therm}$ in Eq. (9); this yields:

$$\varphi_{cath} = -\frac{k_BT_e}{e}\ln\left[\sqrt{\frac{2\pi m_e}{M\left(1+\frac{\pi r_{De}}{2\lambda_{i-mfp}}\right)}}\left(1-\frac{j}{j_i}\right)\right]. \quad (21)$$

This equation has been obtain in [21].

*2B. Anode sheath and sheath at floating wall*

Now let us consider the case of the anode where a refractory metal insert bonded to a high thermal conductivity metal, such as copper, to prevent its erosion. Such anodes are used in high-current pulsed Gerdien arc lamps. The temperature of the insert in theses cathodes can be high enough that the thermionic electron current becomes significant. Because in the case of the anode the total current is directed from the wall to the plasma while in the case of the cathode it is directed from the plasma to the wall, an equation for the total current density at the anode can be obtained from Eq. (3) by simply adding a minus in front of $j$:

$$-j = j_i - j_{e-plasma} + j_{e-therm}. \qquad (22)$$

Thus, substituting $-j$ instead of $j$ into Eqs. (9) and (20) we obtain the anode sheath potential drop, Fig. 1,

$$\varphi_{anode} = -\frac{k_B T_e}{e} \ln\left[\sqrt{\frac{2\pi m_e}{M\left(1+\frac{\pi r_{De}}{2\lambda_{ch-exch}}\right)}}\left(1+\frac{j+j_{e-therm}}{j_i}\right)\right]. \qquad (23)$$

and the following system of equations for determination $\varphi_{anode}^{critical}$ and $j_{e-therm}^{critical}$ in the case of virtual anode:

$$exp\left(-\frac{e\varphi_{anode}^{critical}}{k_B T_e}\right) + 4\alpha\left(\sqrt{1+\frac{e\varphi_{anode}^{critical}}{k_B T_e}\frac{1}{\alpha}}-1\right) - \frac{4j_{e-therm}^{critical}}{en_p\sqrt{\frac{2k_B T_e}{m_e}}}\sqrt{\frac{e\varphi_{anode}^{critical}}{k_B T_e}} - 1 = 0, \qquad (24)$$

$$\varphi_{anode}^{critical} = -\frac{k_B T_e}{e} \ln\left[\sqrt{\frac{2\pi m_e}{M\left(1+\frac{\pi r_{De}}{2\lambda_{i-mfp}}\right)}}\left(1+\frac{j+j_{e-therm}^{critical}}{j_i}\right)\right] \qquad (25)$$

In the case of cold anode (no thermionic emission or erosion of the wall) the anode sheath potential drop can be obtained by dropping $j_{e-therm}$ in Eq. (23), this yields [21]:

$$\varphi_{anode} = -\frac{k_B T_e}{e} \ln\left[\sqrt{\frac{2\pi m_e}{M\cdot\left(1+\frac{\pi \cdot r_{De}}{2\cdot\lambda_{ch-exch}}\right)}}\left(1+\frac{j}{j_i}\right)\right]. \qquad (26)$$

It should be stressed that the anode fall, a quasineutral near anode boundary layer with the thickness of a few electron mean free paths [25], is much larger than $r_{De}$, the thickness of the anode sheath. The anode fall can completely disappear or even become positive as it has been observed in the case of large current densities in weakly ionized plasmas [25].

In the case of cold floating walls, $j = 0$, the floating sheath potential drop [21] can be obtained from Eq. (26) by dropping $j$; this yields:

$$\varphi_{float} = -\frac{k_B T_e}{e} \ln\left[\sqrt{\frac{2\pi m_e}{M\left(1+\frac{\pi r_{De}}{2\lambda_{ch-exch}}\right)}}\right]. \tag{27}$$

*2C. Boundary conditions at the walls*

Let us first consider the case of the arc with a thermionic electron emitting cathode and a cold anode with no thermionic electron emission, Fig. 3. Boundary conditions at the floating walls, contours D-E-F in Fig. 3, is $j = 0$; which may be written as:

$$\frac{\partial \varphi}{\partial n} = 0, \tag{28}$$

where $\partial/\partial n$ is the space derivative normal to the wall. Because the current through the cathode thermionic insert housing, contour B-C in Fig. 3, is usually negligibly small in comparison to the current through the thermionic cathode insert, it can be taken as zero and, therefore, yielding the boundary condition given by Eq. (28) at contour B-C as well. The boundary condition at the thermionic cathode has to be taken at the plasma-sheath interface, Fig. 1, and is

$$\varphi = \varphi_{cath} \tag{29}$$

where $\varphi_{cath}$ is determined by numerically solving the set of Eqs. (7) - (9), and (18), the case of no virtual cathode, or the set of Eqs. (19) and (20), the case of virtual cathode. In Eq. (29) the potential of the thermionic cathode surface facing the plasma is assumed to be zero. It should be stressed that this boundary condition assumes that the electrical conductivity of cathode material is infinitely large. This is justified because the electrical resistivity of the plasma is significantly larger than the resistivity of the cathode material. The boundary condition at the anode, contour G-H in Fig. 3 can be written as

$$\varphi = \Delta V + \varphi_{anode}, \qquad (30)$$

where $\varphi_{anode}$ is given by Eq. (26), and $\Delta V$ is the cathode - anode voltage drop, Fig. 1. It is worth noting that in the case of a small sheath anode potential drop, this boundary condition can be simplified to

$$\varphi = \Delta V, \qquad (31)$$

which has been used in all previous hydrodynamic modeling of high-pressure arcs, see [1 - 18] and references therein.

It should be noted also that if a constant current power supply is used, the voltage $\Delta V$ in the simulation has to be iterated until the calculated total arc current is equal to the current setting of the power supply.

In the case of the anode with thermionic electron emitter, Fig.4, the boundary conditions at the anode are similar to the boundary conditions at the cathode with thermionic electron emitter. The boundary conditions at contour G-H, Fig. 4, is given by Eq. (28) and at contour H-I is by Eq. (30), where $\varphi_{anode}$ is determined by numerically solving the set of Eqs. (7), (8), (23), and (18), the case of no virtual anode, or the set of Eqs. (24) and (25), the case of virtual anode.

In the case of a cold biased electrode, the boundary condition at the wall is

$$\varphi = \varphi_{bias} + \varphi_{electrode} \qquad (32)$$

where $\varphi_{bias}$ is the biased voltage of the electrode and $\varphi_{electrode}$ is set to either $\varphi_{anode}$ or $\varphi_{cath}$, Eq. (26) or (21) depending on weather the electrode is a biased anode or a biased cathode.

**III. Boundary conditions at the wall for the electron and heavy particles energy equations**

Let us first consider the case of the cathode with thermionic electron emission. The enthalpy flux from the plasma to the wall due to the charged particles that reach the wall can be written as [19]:

$$q_{charged}^{particles} = e n_p V_s \left( I_{ioniz} + \varphi_{cath} + \frac{M V_s^2}{2e} \right) + 2 k_B T_e n_p \exp\left( -\frac{e \varphi_{cath}}{k_B T_e} \right) \sqrt{\frac{k_B T_e}{2 \pi m_e}}, \qquad (33)$$

where $I_{ioniz}$ is the ionization potential of the working gas. Eq. (33) assumes that all ions incoming into the sheath reach the wall, recombine there with electrons, and come back to the plasma as neutrals where they are immediately ionized by electrons. The first term on the right-hand side of Eq. (33), describes the heat flux to the wall due to the recombination process plus the kinetic energy flux that ions bring to the wall (directly, or by fast atoms created in the charge exchange process), and the second term describes the heat flux that electrons bring to the wall. Since we assumes $k_B T_h \ll e \varphi_{cath}$, in Eq. (33) we have neglected the ion thermal heat flux to the wall. It is worth noting that the third term in the first brackets is the kinetic energy of an ion entering the sheath.

The thermionic electrons accelerated in the sheath carry their enthalpy to the plasma

$$q_{electrons}^{thermion} = j_{e-therm} \varphi_{cath}, \qquad (34)$$

where $j_{e-therm}$ and $\varphi_{cath}$ are calculated in Section 2A. In Eq. (33), as in Eq. (34), we have neglected the thermal energy flux that the thermionic electrons bring to the plasma. Because the electron-electron energy transfer collision frequency is much larger than the electron-heavy particle energy exchange rate, the thermionic electrons transfer the energy they gained in the sheath only to the "plasma" electrons, not to the heavy particles. Because the heat flux to the wall $q_{charged}^{particles}$, Eq. (33), is due to the change of energy of electrons only, not heavy particles [19], setting $q_{charged}^{particles} - q_{electrons}^{thermion}$ equal to the electron enthalpy flux from the plasma to the sheath we obtain the following boundary condition for electron energy equation at the cathode:

$$-\kappa_e \frac{\partial T_e}{\partial n} - \frac{j}{e}\left(\frac{5}{2} k_B T_e\right) = e n_p V_s \left( I_{ioniz} + \varphi_{cath} + \frac{M V_s^2}{2e} \right) +$$

$$+2k_B T_e n_p \exp\left(-\frac{e\varphi_{cath}}{k_B T_e}\right)\sqrt{\frac{k_B T_e}{2\pi m_e}} - j_{e-therm}\varphi_{cath}, \tag{35}$$

where the second term on the left hand side of Eq. (35) describes the electron translation enthalpy flux which is directed from the cathode [16]; $\kappa_e$ is the electron thermal conduction coefficient. Thus, in the case where the temperature of the wall is given, the Dirichlet boundary condition, a set of boundary conditions for the electron and heavy particle energy equations at the cathode can be written as:

$$\frac{\partial T_e}{\partial n} = -\frac{en_p V_s}{\kappa_e}\left(I_{ioniz} + \varphi_{cath} + \frac{MV_s^2}{2e}\right) - \frac{2k_B T_e n_p}{\kappa_e}\exp\left(-\frac{e\varphi_{cath}}{k_B T_e}\right)\sqrt{\frac{k_B T_e}{2\pi m_e}} -$$
$$-\frac{j}{e\kappa_e}\left(\frac{5}{2}k_B T_e\right) + \frac{1}{\kappa_e}j_{e-therm}\varphi_{cath}, \tag{36}$$

$$T_h = T_{sur}. \tag{37}$$

It has to be stressed that the total thermal heat flux to the wall due to charged particles coming from the plasma is

$$Q_{charged}^{particles} = q_{charged}^{particles} - j(\varphi_{work-func} - \Delta\varphi_{Schot}). \tag{38}$$

The second term on the right hand side of Eq. (38) describes the energy flux that the cathode loses because the electrons from the cathode leave the electrode to recombine with the plasma ions at the electrode surface and create the thermionic electron current to provide the total current density $j$.

In the case of the Neumann boundary condition, a set of boundary conditions for the electron and heavy particles energy equations at the cathode can be written as:

$$\frac{\partial T_e}{\partial n} = -\frac{en_p V_s}{\kappa_e}\left(I_{ioniz} + \varphi_{cath} + \frac{MV_s^2}{2e}\right) - \frac{2k_B T_e n_p}{\kappa_e}\exp\left(-\frac{e\varphi_{cath}}{k_B T_e}\right)\sqrt{\frac{k_B T_e}{2\pi m_e}} -$$
$$-\frac{j}{e\kappa_e}\left(\frac{5}{2}k_B T_e\right) + \frac{1}{\kappa_e}j_{e-therm}\varphi_{cath}, \tag{39}$$

$$-\kappa_w \frac{\partial T_w}{\partial n} = en_p V_s \left(I_{ioniz} + \varphi_{cath} + \frac{MV_s^2}{2e}\right) + 2k_B T_e n_p \exp\left(-\frac{e\varphi_{cath}}{k_B T_e}\right) \sqrt{\frac{k_B T_e}{2\pi m_e}} -$$
$$-j(\varphi_{work-func} - \Delta\varphi_{Schot}) - \kappa_h \frac{\partial T_h}{\partial n} + Rad, \tag{40}$$

where $-\kappa_w \partial T_w/\partial n$ is the heat flux in the wall, $-\kappa_h \partial T_h/\partial n$ is the heat flux of heavy particles to the wall, $Rad$ is the net radiation heat flux of the wall, and index $w$ corresponds to wall. In this paper we are not specifying $Rad$. As one can see in Eq. (40), we used the total thermal conduction coefficient of heavy particles $\kappa_h$ instead of $\kappa_n$ - the thermal conduction coefficient of neutrals. This is favorable since available databases for plasma transport properties provide the total thermal conduction coefficients for heavy particles, $\kappa_h$, without dividing it into $\kappa_i$ and $\kappa_n$, [26 - 27]. However, since $\kappa_e \gg \kappa_i$ using $\kappa_h$ instead of $\kappa_n$ should not lead to significant errors in simulating the heat transfer from the plasma to the wall.

Now let us consider the case where the electrode is the anode with thermionic electron emission. Following Section 2B, putting minus in front of $j$ in Eqs. (36), (39) and (40) we obtain the two sets of boundary conditions for the electron and heavy particles energy equations at the cathode:

$$\frac{\partial T_e}{\partial n} = -\frac{en_p V_s}{\kappa_e}\left(I_{ioniz} + \varphi_{anode} + \frac{MV_s^2}{2e}\right) - \frac{2k_B T_e n_p}{\kappa_e} \exp\left(-\frac{e\varphi_{anode}}{k_B T_e}\right) \sqrt{\frac{k_B T_e}{2\pi m_e}} +$$
$$+\frac{j}{e\kappa_e}\left(\frac{5}{2} k_B T_e\right) + \frac{1}{\kappa_e} j_{e-therm} \varphi_{anode}, \tag{41}$$

$$T_h = T_{sur}, \tag{42}$$

and

$$\frac{\partial T_e}{\partial n} = -\frac{en_p V_s}{\kappa_e}\left(I_{ioniz} + \varphi_{anode} + \frac{MV_s^2}{2e}\right) - \frac{2k_B T_e n_p}{\kappa_e} \exp\left(-\frac{e\varphi_{anode}}{k_B T_e}\right) \sqrt{\frac{k_B T_e}{2\pi m_e}} +$$
$$+\frac{j}{e\kappa_e}\left(\frac{5}{2} k_B T_e\right) + \frac{1}{\kappa_e} j_{e-therm} \varphi_{anode}, \tag{43}$$

$$-\kappa_w \frac{\partial T_w}{\partial n} = en_p V_s \left(I_{ioniz} + \varphi_{anode} + \frac{MV_s^2}{2e}\right) + 2k_B T_e n_p \exp\left(-\frac{e\varphi_{anode}}{k_B T_e}\right) \sqrt{\frac{k_B T_e}{2\pi m_e}} +$$

$$+j(\varphi_{work-func} - \Delta\varphi_{Schot}) - \kappa_h \frac{\partial T_h}{\partial n} + Rad \ . \tag{44}$$

where $\varphi_{anode}$ and $j_{e-therm}$ are calculated in Section 2B. The first set, Eqs. (41) and (42), corresponds to the case where the temperature of the wall is given, the Dirichlet boundary conditions, and the second one, Eqs. (43) and (44), where the heat flux to the wall is given, represents the Neumann boundary conditions.

In the case of electrodes with no thermionic emission the obtained boundary conditions for the electron and heavy particle energy equations at the electrodes reduce to the boundary conditions for cold biased wall [19] by dropping $j_{e-therm}$ in Eqs. (36), (39), (41), and (43). Putting $j = j_{e-therm} = 0$ and using $\varphi_{float}$ instead of $\varphi_{catht}$ in Eqs. (36), (39), and (40) we obtain the boundary conditions for the energy equations at the floating wall [19].

## IV. An algorithm of implementation of boundary conditions

Since the suggested boundary conditions for $\varphi$, $T_e$, $T_h$, and $T_w$ link the Poisson equation for the electrical potential with the energy equations for electrons and heavy particles in the plasma and the heat transfer equation in the wall their implementation in a numerical algorithm is not trivial. Therefore, we would like to suggest an algorithm of how these boundary conditions might be implemented. In two temperature thermal plasma approximation the variables are: three thermodynamics parameter of the plasma (usually ($T_e$, $T_h$, $P$) or the enthalpies of electrons and heavy particles and the mass density of the plasma), the flow velocity of the plasma $\vec{u}$, the electric potential $\varphi$, and the wall temperature $T_w$. In the suggested algorithm below we consider the following parameters $T_e$, $T_h$, $P$, $\vec{u}$, $\varphi$, and $T_w$; this algorithm can be applied to other sets of thermodynamic parameters of the plasma as well.

*Algorithm*

Step 1. Let us assume that at iteration step $N$ (in the case of steady state) or at a given time $t$ (in the time dependent case) the following distribution: $T_e$, $T_h$, $P$, $\vec{u}$, $\varphi$, $T_w$, are known.

Step 2. Calculate the current density distribution in the plasma $j$ by using the following equation: $j = -\sigma \nabla \varphi$.

Step 3. Go to the next iteration step $N+1$, or to the next time step $t = t + \tau$, where $\tau$ is a time step.

Step 4. Calculate new values of $T_e$, $T_h$, $T_w$ $\vec{u}$, and $P$ by solving the momentum, mass, and energy equations using the boundary conditions at the wall for electron and heavy particle energy equations presented in Section III.

Step 5. For obtained in Step 4 new $T_e$, $T_h$, and $P$ distributions calculate all needed plasma parameters including plasma composition, $r_{De}$, $\lambda_{i-mfp}$, $j_i$, $V_s$, $\sigma$, and others used in Sections II and III.

Step 6. Calculate new $j_{e-term}$, $\varphi_{cath}$, $\varphi_{anode}$, and $E_{surf}$, Sections II.

Step 7. Solve Eq. (1) for $\varphi$ using the boundary conditions at the walls presented in Section II.

As one can see Step 7 ends the algorithm, see Step 1.

## V. Virtual cathode, numerical results

In this Section we illustrate the formation of the virtual cathode, with thermionic electron emission, for singly ionized argon plasma. In our simulation the plasma pressure $P$ was $4 \cdot 10^5$ Pa and the current density in the cathode spot $j$ is $3.3 \cdot 10^8$ A/m$^2$ which are typical parameters for modeling 200A plasma cutting torch [10]. Following [16], the plasma composition is determined by solving the Saha equation with $T_e$ at given plasma pressure $P$ and given temperature of heavy particles $T_h$:

$$\frac{n_e^2}{n_n} = 2\left(\frac{2\pi m_e k_B T_e}{h^2}\right)^{3/2} \frac{Q_{Ar^+}(T_e)}{Q_{Ar}(T_e)} exp\left(-\frac{eI_{ioniz}}{k_B T_e}\right) = 2.89 \times 10^{22} T_e^{3/2} exp\left(-\frac{1.827 \times 10^5}{T_e}\right), \quad (45)$$

$$P = k_B(n_e + n_n)T_h + k_B n_e T_e, \quad (46)$$

where $n_n$ is the number density of neutral argon, $n_e$ is the electron number density which is equal to $n_p$, and $Q_{Ar^+}(T_e)$ and $Q_{Ar}(T_e)$ are the statistical sums of partition functions of argon ions and argon neutral atoms respectively. Two assumptions were made in Eq. (45) and (46): (1) the contributions of the excited states to the statistical sums $Q_{Ar^+}$ and $Q_{Ar}$ are less than 5 percent [28], and therefore, have been neglected in Eq. (45); (2) because the number densities of multi-charged ions are many orders of magnitude smaller than the number density of singly ionized argon, multi-charged ions are ignored in this model. In this simulation we chose the temperature of heavy particles to be 3500K (which in our model is equal to the surface temperature of the thermionic cathode wall $T_{sur}$) and $T_e = 9000$ K.

The electrical field at the cathode surface and the cathode sheath potential drop vs. the thermionic electron density are shown in Figs. 5 and 6. In these simulations we used $\sigma_{i,n} = 1.18 \cdot 10^{-18}$ m$^2$ [29] which is the total $Ar^+ - Ar$ momentum transfer cross section. As one can see from Fig, 6, the electric field at the cathode reaches zero at $j_{e-therm} = j_{e-therm}^{critical}$. Further increase in $j_{e-therm}$, $j_{e-therm} > j_{e-therm}^{critical}$, lead to negative values in the term under the square root in Eq. (15). In other words, for the selected parameters of the plasma and the surface temperature of the wall there exists no solution for $j_{e-therm} > j_{e-therm}^{critical}$. Fig. 7 shows the electrical field distribution in the sheath, $e\varphi/k_B T_e$, vs. $x/L_{cath}^{sheath}$ for different $j_{e-therm}$; $L_{cath}^{sheath}$ is the thickness of the cathode sheath. In this figure the electrical potential at the plasma sheath was chosen to be zero as in Fig. 1. As expected, with an increase in the $j_{e-therm}$ the value of $\varphi_{cath}$ decreases as does the thickness of the sheath $L_{cath}^{sheath}$.

Plasma sheath model [20, 21] assumes that $E_{plasma}$, the electrical field in the plasma at the plasma-sheath interface, has to be much smaller than $k_B T_e/er_{De}$, the electrical field in the sheath at the sheath side, Eq. (12). This assumption can be shown to be valid for this simulation. The plasma conductivity at the sheath can be written as:

$$\sigma = \frac{e^2 n_e}{\nu_{e,i}^{tr} + \nu_{e,n}^{tr}}, \qquad (47)$$

where

$$\nu_{e,i}^{tr} = 4.93 \times 10^{-6} \frac{\Lambda n_e}{T_e^{3/2}}, \qquad \Lambda = 18.7 - \ln\left(\frac{n_e^{1/2}}{T_e^{5/4}}\right), \qquad (48)$$

$$\nu_{e,n}^{tr} = n_n \sigma_{e,n} \sqrt{\frac{k_B T_e}{m_e}}; \qquad (49)$$

$\nu_{e,i}^{tr}$ was taken from [30]. In our model we have used $\sigma_{e,n} = 2 \times 10^{-20}$ m$^2$ which was extracted from data in [31]. Substituting $n_e = 1.76 \times 10^{22}$ m$^3$ and $n_n = 8.22 \times 10^{24}$ m$^3$ (calculated using Eqs. (45) and (46)) into Eqs. (48) and (49) we obtain the value of $\sigma = 6.32 \times 10^2$ A/Vm and $E_{plasma} = j/\sigma = 6.01 \times 10^5$ V/m. As one can see the obtained $E_{plasma}$ is 26 times smaller than $k_B T_e / e r_{De}$. Thus, the assumption used is well satisfied.

As it was mentioned in Section II, the model assumes that friction force between ions and electrons in the sheath is much smaller than between ions and neutrals and, therefore, can be neglected. This assumption can be shown to be valid for this simulation as well. The ratio of ion-electron friction force to the ion-neutral friction force in the sheath can be estimated as:

$$Friction_{i,n}^{i,e} \sim \frac{m_e \cdot (\nu_{i,e-plasma}^{tr} + \nu_{i,e-therm}^{tr})}{M \nu_{i,n}^{tr}}, \tag{50}$$

where $\nu_{i,e-plasma}^{tr}$ and $\nu_{i,e-therm}^{tr}$ are the collision frequensies of an ion in the sheath with the plasma electrons and the thermionic electrons respectively; and $\nu_{i,n}^{tr} \approx n_n \sigma_{i,n} V_s$. Substituting $j_{e-therm}/e\sqrt{2e\varphi_{cath}/m_e}$ and $e\varphi_{cath}/k_B$, the characteristic thermionic electron number density and the characteristic energy of thermionic electons in the sheath, into Eq. (48) instead of $n_e$ and $T_e$, one can estimate $\nu_{i,e-therm}^{tr}$. Substituting in Eq. (48) $n_e = 1.76 \times 10^{22}$ m$^3$ and $T_e = 9000$ K, the number density and the temperature of the electron in front of the sheath, we obtain an estimate for $\nu_{i,e-plasma}^{tr}$. For the range of $j_{e-therm}$ considered in this simulation, $Friction_{i,n}^{i,e}$ reaches its maximum for $j_{e-therm} = j_{e-therm}^{critical}$ (where the sheath potential drop is the smallest and the thermionic electron number density is the largest) and is equal to $7.5 \times 10^{-4}$. Thus, neglecting the friction of ions with electrons in this simulation is appropriate.

The following reviews the validity of assumption (5) where the thermionic electrons pass through the sheath collisionlesly such that $L_{cath}^{sheath} < \lambda_{e-therm-mfp}$, see Eq. (2). The transport thermionic electron mean free path in the sheath can be estimated as

$$\frac{1}{\lambda_{e-therm-mfp}} \approx n_p \cdot \frac{6.43 \times 10^{-10}}{\left(\frac{e\varphi_{cath}}{k_B}\right)^2} \Lambda + n_n \sigma_{e,n}, \tag{51}$$

where the first term on the right hand side in Eq. (51) describes the collisions of a thermionic electron with the plasma electrons and ions, and the second term describes the collisions with the neutrals. Substituting $n_p = 1.76 \times 10^{22}$ m$^{-3}$, $n_n = 8.22 \times 10^{24}$ m$^{-3}$, $\sigma_{e,n} = 2 \times 10^{-20}$ m$^2$, $\Lambda = 5$, along with $\varphi_{cath}$, calculated at a given $j_{e-therm}$ into Eq. (51), we obtain that the ratio of $\lambda_{e-therm-mfp}$ to $L_{cath}^{sheath}$ is larger than 20 in the full range of $j_{e-therm}$ considered in this simulation. Thus, neglecting the collisions of the thermionic electrons in the sheath is appropriate in this example.

Now let us check if $\Delta\varphi/\varphi_{cath} \ll 1$, Eq. (15). As follows from Eq. (15), $\Delta\varphi/\varphi_{cath}$ reaches its maximum for the virtual cathode conditions, where $j_{e-therm}$ is maximum, and $\varphi_{cath}$ is minimum. Substitution these values into Eq. (15) we obtain that $\Delta\varphi/\varphi_{cath} = 0.073$. Thus, neglecting $\Delta\varphi$ in the model is appropriate.

## VI. Model of cathode spot

To demostrate the application of the suggested boundary conditions, they are appled to a zero dimensional model of the cathode spot formed at a tungsten emissive element in argon plasma. In this model we assume: (1) the surface temperature of the tungsten cathode, $T_{sur}$, is constant; (2) the heat loss from the plasma to the cathode is compensated by the energy that the thermionic electrons bring to the plasma, (3) no electron thermal conduction in the plasma, $\partial T_e/\partial n = 0$ at the sheath-plasma interface. (4) In the model, as in Section V, we assume that the argon plasma is singly ionized and calculate the plasma compositions using Eqs. (45) and (46). The model neglects the 2D edge efects of the spot because the thickness of the cathode sheath, $L_{cath}^{sheath}$, is assumed to be much smaller than the diameter of the cathode spot. Therefore, the model gives the right order of magnitude for the cathode spot current density $j$.

As follows from assumptions (2) and (3), the Dirichlet boundary conditions at the cathode surface, Eqs. (36) and (37), reduces to the following equation for $j$:

$$j_{e-therm}\varphi_{cath} =$$
$$= en_p V_s \left(I_{ioniz} + \varphi_{cath} + \frac{MV_s^2}{2e}\right) + 2k_B T_e n_p exp\left(-\frac{e\varphi_{cath}}{k_B T_e}\right)\sqrt{\frac{k_B T_e}{2\pi m_e}} + j\left(\frac{5k_B T_e}{2e}\right), \quad (52)$$

where $\varphi_{cath}$ given by Eq. (9) and $j_{e-therm}$ by Eq. (7). Thus, solving the set of Eqs. (52), (7), (8), (9) and (18) at given $T_e$, $T_{sur} = T_h$, and $P$ with plasma composition calculated by Eqs. (45) and (46), we obtain $j$, $j_{e-therm}$, $e\varphi_{cath}$, $E_{sur}$, and $\Delta\varphi_{Schot}$. The result of the simulations are presented in Figs. (8) - (11). In this simulation we used $A = 6 \cdot 10^5$ A/(m²·K²), $\varphi_{work-func} = 4.54$ eV (the Richardson parameters of Tungsten), $T_{sur} = 3800$ K, and $P = 4 \cdot 10^5$ Pa.

As follows from Fig. 8, in the selected range of the total current densities, $j$, the thermionic electron current density, $j_{e-therm}$, increases more than two time and is equal to 8.2·10⁶ A/m2 for $E_{sur} = 0$. Thus, we have demonstrated that taking into account the Schottky effect is very important in modeling the cathode spot. Moreover, because $\Delta\varphi_{Schot}$ varies significantly with the parameters of the cathode spot (total cathode current density $j$), using incorrect values of $\Delta\varphi_{Schot}$ in the model may leads to misleading results. As follows from Figs. 9 and 10, the electron temperature of the plasma at the sheath-plasma interface and the cathode sheath potential voltage drop increase with an increase in the cathode spot current density, as expected. As shown in Fig. 11, for small sheath current densities the sheath is collisional, $r_{De} > \lambda_{i-mfp}$. With an increase in $j$ the plasma electron temperature increases, the plasma becomes more ionized, $r_{De}$ decreases, and $\lambda_{i-mfp}$ increases leading to a decrease in the collision parameter $1 + 2\pi r_{De}/2\lambda_{i-mfp}$, see for example Eq. (5).

It is worth noting that for high cathode spot current density, where $e\varphi_{cath}/k_B T_e \gg 1$, the plasma "electron" current density in the sheath, $j_{e-plasma}$, is very small, and, therefore, can be dropped in Eq. (3),

$$j = j_i + j_{e-therm}, \tag{53}$$

that leads to reducing Eq. (52) and (18) to the following forms:

$$j_{e-therm}\varphi_{cath} = en_p V_s \left(I_{ioniz} + \varphi_{cath} + \frac{MV_s^2}{2e}\right) + j\left(\frac{5k_B T_e}{2e}\right), \tag{54}$$

$$E_{sur} = \frac{k_B T_e}{er_{De}}\left(4\alpha\left(\sqrt{1 + \frac{e\varphi_{cath}}{k_B T_e}\frac{1}{\alpha}} - 1\right) - 4\beta\sqrt{\frac{e\varphi_{cath}}{k_B T_e}} - 1\right)^{1/2}. \tag{55}$$

Substituting $j$ from Eq. (53) into Eq. (54) and $\alpha$ and $\beta$ from Eq. (14) into Eq. (55), Eqs. (54) and Eq. (55) can be reduced to the following forms:

$$\varphi_{cath} = \frac{I_{ioniz} + \frac{k_B T_e}{2e}\left(1+\frac{\pi r_{De}}{2\lambda_{i-mfp}}\right)^{-1} + \left(\frac{j_{e-therm}}{j_i}+1\right)\frac{5k_B T_e}{2e}}{\frac{j_{e-therm}}{j_i}-1}, \quad (56)$$

$$E_{sur} = \frac{k_B T_e}{e r_{De}}\left(\frac{2}{\left(1+\frac{\pi \cdot r_{De}}{2\cdot\lambda_{i-mfp}}\right)}\left(\sqrt{1+\frac{2e\varphi_{cath}}{k_B T_e}\left(1+\frac{\pi \cdot r_{De}}{2\cdot\lambda_{i-mfp}}\right)}-1\right) - \frac{j_{e-therm}}{en_p k_B T_e}\sqrt{8m_e e\varphi_{cath}}-1\right)^{1/2} \quad (57)$$

where $j_i$ is given by Eq. (10). Solving the set of Eqs. (56), (57), (7) and (8) we determine $E_{sur}$, $\Delta\varphi_{Schot}$, $j_{e-therm}$, and $\varphi_{cath}$. Then, substituting obtained $j_{e-therm}$ into Eq. (53) we obtain $j$. In our simulation we used this approximation for $e\varphi_{cath}/k_B T_e > 11.3$, Fig. 8, where the ratio of $j_{e-plasma}$ to $j_i$ was smaller than 0.002. As one can see from Eq. (56), when the ion current density, $j_i$, reaches the thermionic electron current density, $j_{e-therm}$, the $\varphi_{cath} \to \infty$ and $j$ reaches its maximum value of $2j_{e-therm}$, see Eq. (53). This result is physically reasonable: for $\varphi_{cath} \to \infty$ the heat flux from the plasma to the wall is $j_i \varphi_{cath}$ which in this model has to be balanced by the energy flux that the thermionic electrons bring to the plasma, $j_{e-therm}\varphi_{cath}$, leading to $j_{e-therm} = j_i$.

In the case where the thermal electron heat flux at the plasma-sheath interface is not zero, $\partial T_e/\partial n < 0$, (not considered in this cathode spot model) the energy flux from the plasma to the cathode is balanced not only by the energy that the thermionic electrons bring to the plasma but also by the electron thermal heat flux coming from plasma to the sheath, $-\kappa_e \partial T_e/\partial n$, Eq. (36). This leads to a decrease in the required $T_e$ and correspondingly in $\varphi_{cath}$ to maintain the arc compared to the case considered in this model where $\partial T_e/\partial n = 0$.

It is interesting to note that the formulated boundary conditions predicts the existence of an arc even in the case of zero thermionic electron emission where the heat flux to the electrode is balanced by the electron thermal heat flux, $-\kappa_e \partial T_e/\partial n$, directed from the plasma to the plasma-sheath interface. This regime likely to exist in very low current density arcs with very well cooled electrodes to prevent the evaporation of the electrode.

It should be noted that the presented model of the cathode spot is similar to model [32]. However, there are some significant differences between the models: (1) In [32] the temperature of heavy particle in the sheath is assumed to be much larger than the temperature of the cathode surface and equal to 10000K. In our model, $T_h$ is equal to the temperature of the cathode surface, $T_{sur}$. (2) The model in [32] assumes that the sheath is collisionless. As shown in Fig. 11, this assumption is not always valid. (3) Also, [32] neglects the contribution of thermionic electron number density in the Poisson equation, the third term in the brackets in Eq. (11) is absent in their formulation. Therefore, [32] will not predict the virtual cathode and overestimates $\Delta\varphi_{Schot}$. (4) In our formulation the plasma sheath is considered as the plasma-wall interface where the plasma parameters at the sheath are calculated directly by a two temperature thermal plasma model. However, [32] uses the ionization layer, a layer where plasma is not in chemical equilibrium, as an intermediate layer between the sheath and plasma. As follows from [32], the voltage drop across the ionization layer is much smaller than the voltage drop in the sheath. (5) In our model we neglect the contributions of $T_h$ in the plasma sound speed and in the energy balance equations across the sheath while [32] doesn't. Because in our model $T_h = T_{sur} \ll T_e$, such a simplification should not lead to significant variations.

We have checked all assumptions made in the model and found that all of them are very well satisfied for all the range of parameters considered.

**VII. Conclusions**

A new boundary conditions at the electrodes with thermionic electron emission for two temperature thermal arc models have been derived. The obtained boundary conditions take into account the plasma sheath formed at the walls. In terms of two temperature modeling thermal arcs, in the current work, the current profile on the cathode surface is no longer imposed, as in some previous models, but rather calculated. This makes the current model of cathode-plasma interaction self-consistent. The obtained boundary conditions reduce to the boundary conditions for cold floating walls and cold biased electrode (where thermionic electron emission or evaporation of the wall can be neglected) presented in [19]. We have also obtained the boundary condition for the electrical potential at the electrode for the case where thermionic electron current densities emitted from the wall are large enough that a virtual electrode is formed at the walls. We have demonstrated the formation of virtual cathode for the case of a singly ionized argon plasma at $5 \cdot 10^5$ Pa plasma pressure with current densities of $3.8 \cdot 10^8$ A/m², which are typical for modeling 200 A plasma cutting torches [10].

We apply the obtained boundary conditions for a zero dimension model of the cathode spot in which the electron thermal heat flux from the plasma to the wall is zero, $\partial T_e/\partial n = 0$ at the plasma-sheath interface. The obtained results confirmed the significance of incorporating the Schottky correlation factor for calculating thermionic electron emission currents in modeling plasma cutting arcs.

An algorithm of implementation of these boundary conditions in a two temperature thermal plasma model is suggested.

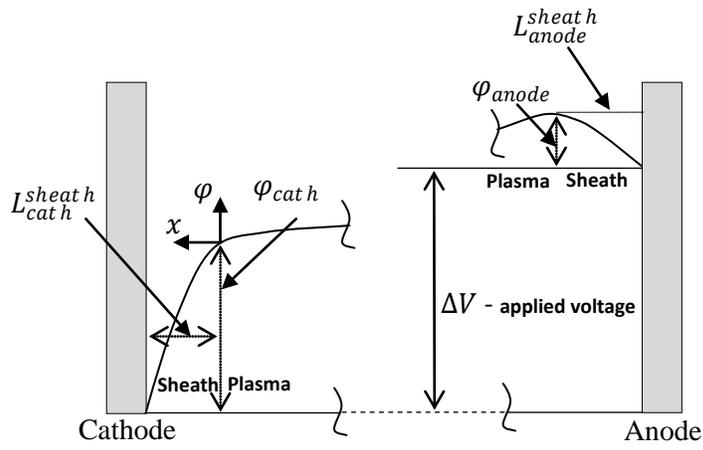

Fig. 1. Schematic of the electrical potential distribution in the arc.

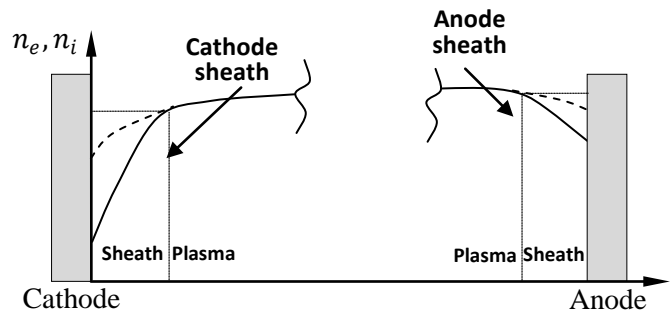

Fig. 2. Schematic of the $n_e$ (solid line) and $n_i$ (dashed line) distributions in the arc.

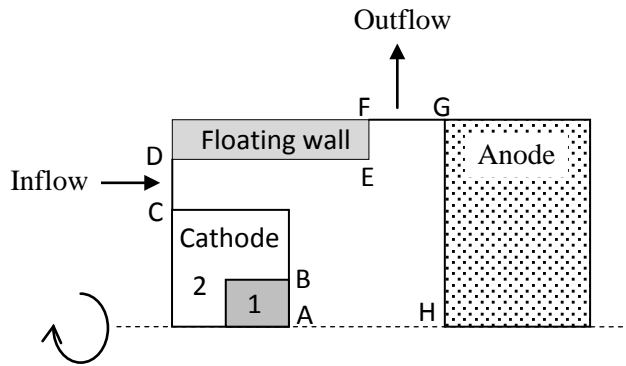

Fig. 3. Schematic of a constricted arc setting with thermionic cathode insert: **1** - the thermionic cathode insert, **2** - the housing of the cathode insert.

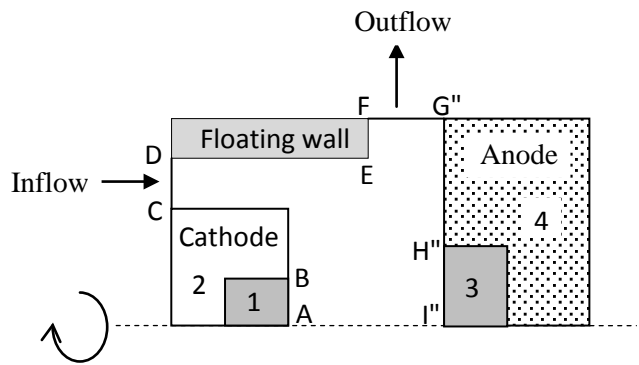

Fig. 4. Schematic of the constricted arc setting with thermionic cathode and anode inserts: **1** - the thermionic cathode insert, **2** - the housing of the cathode insert, **3** - the thermionic anode insert, **4** - the housing of the anode insert.

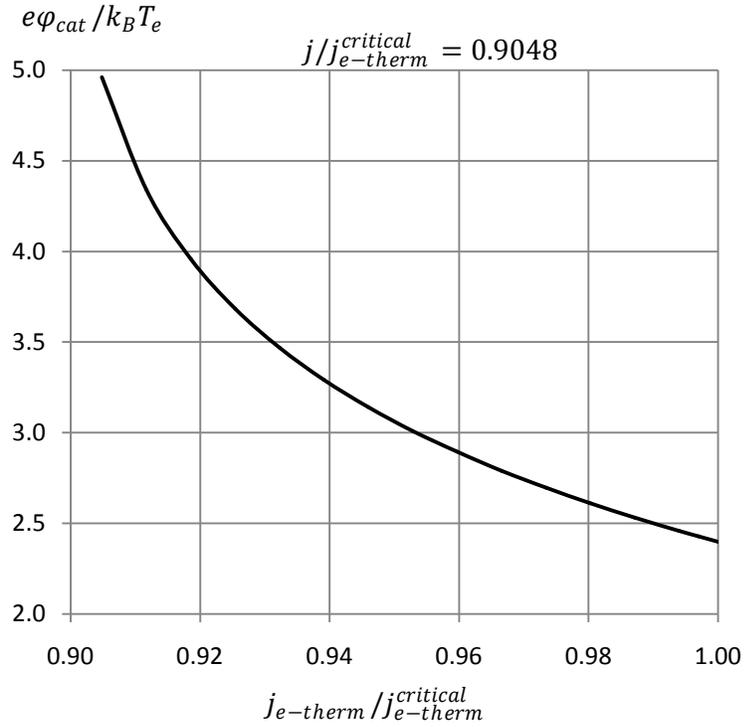

Fig. 5. The cathode sheath potential drop vs. thermionic electron current density in the singly ionized argon plasma. Parameters of the discharge are:
$T_e = 9000\ K, P = 4 \cdot 10^5\ Pa, j = 3.3 \cdot 10^8 A/m^2$.

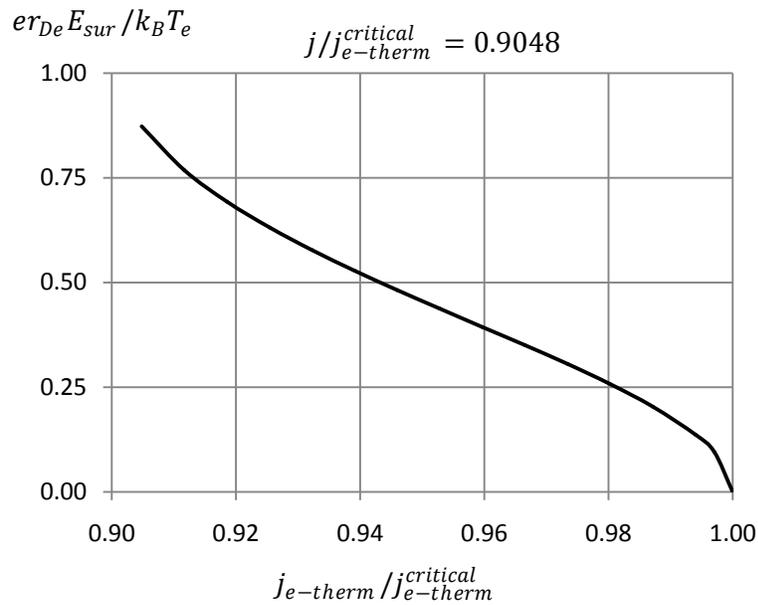

Fig. 6. The electrical field at the cathode surface, $E_{sur}$, vs. thermionic electron current density in the singly ionized argon plasma. Parameters of the discharge are:
$T_e = 9000\ K, P = 4 \cdot 10^5\ Pa, j = 3.3 \cdot 10^8 A/m^2$.

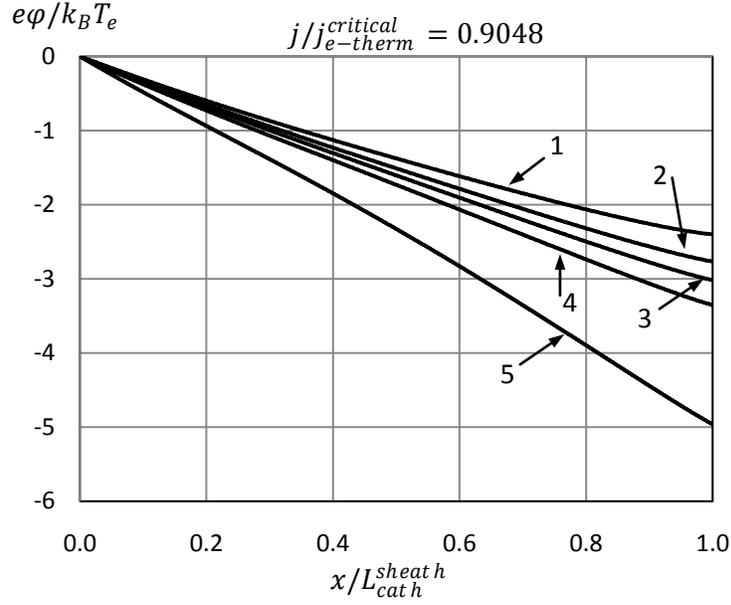

Fig.7. The electrical field distribution for different $j_{e-therm}$; $x = 0$ corresponds to the plasma-sheath boundary and $x/L_{sheath} = 1$ corresponds to the cathode surface Parameters of the discharge are: $T_e = 9000\ K, P = 4 \cdot 10^5\ Pa, j = 3.3 \cdot 10^8 A/m^2$.

1 - $j_{e-therm} = j_{e-therm}^{critical}$, $L_{sheath} = 3.2 \cdot rDe$;
2 - $j_{e-therm} = j - (j_{e-therm}^{critical} - j)/3$, $L_{cath}^{sheath} = 3.46 \cdot rDe$
3 - $j_{e-therm} = j - (j_{e-therm}^{critical} - j)/2$, $L_{cath}^{sheath} = 3.65 \cdot rDe$
4 - $j_{e-therm} = j - 2(j_{e-therm}^{critical} - j)/3$, $L_{cath}^{sheath} = 3.90 \cdot rDe$
5 - $j_{e-therm} = j$, $L_{cath}^{sheath} = 5.00 \cdot rDe$

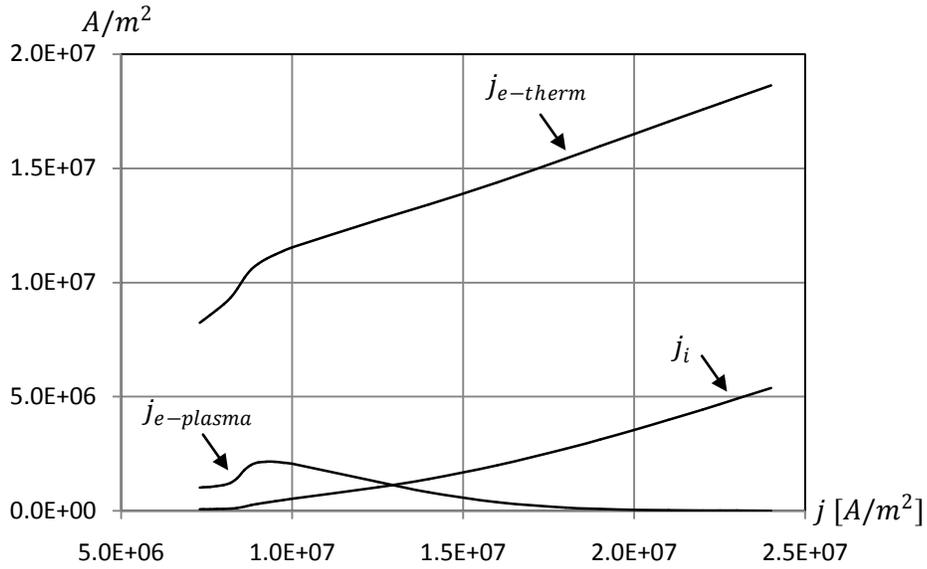

Fig. 8. Model of the cathode spot: $j_{e-therm}$ - thermionic electron current density, $j_i$ - ion current density, $j_{e-plasma}$ - electron current density from the plasma in the sheath; $j$ - the total cathode current density in the sheath. $j_{e-therm}(3800K, E_{sur} = 0) = 8.2 \cdot 10^6$ A/m².

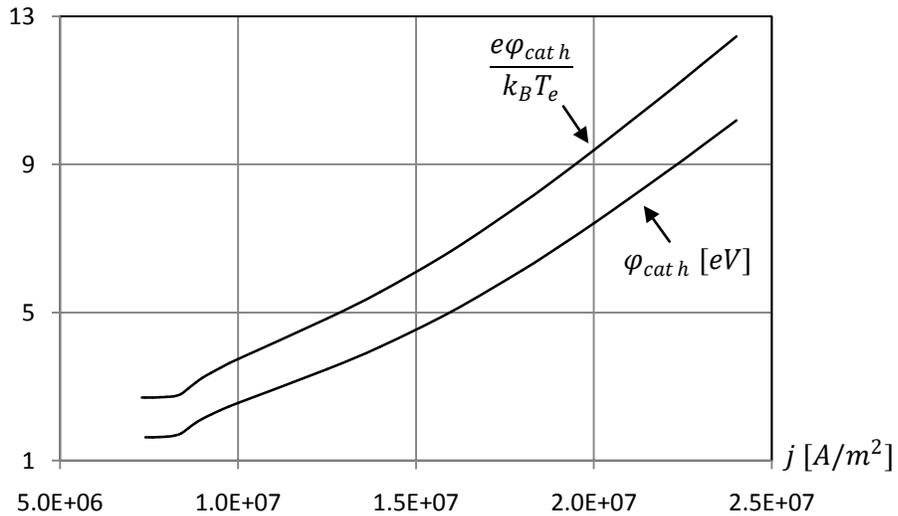

Fig. 9. Model of the cathode spot: $j_{e-therm}$ - thermionic electron current density, $j_i$ - ion current density, $j_{e-plasma}$ - electron current density from the plasma in the sheath; $j$ - the total cathode current density in the sheath. $j_{e-therm}(3800K, E_{sur} = 0) = 8.2 \cdot 10^6$ A/m².

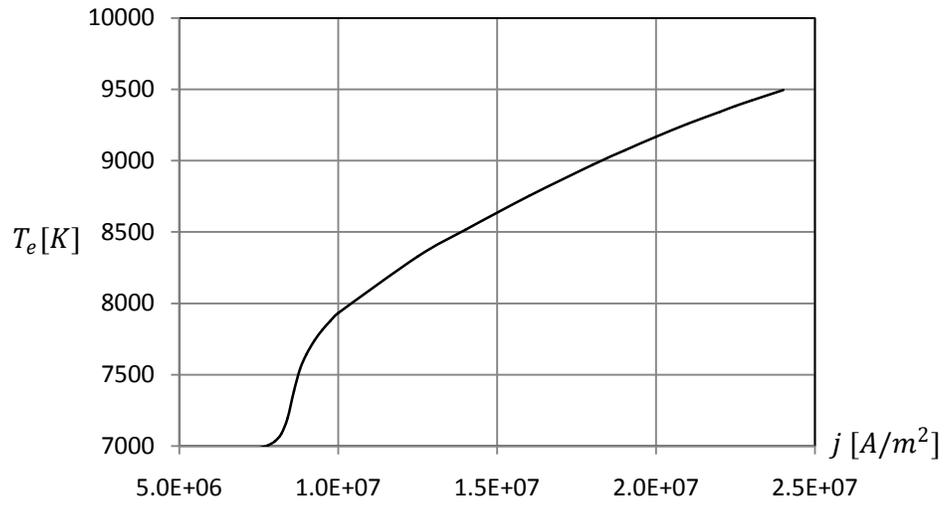

Fig. 10. Model of the cathode spot: $T_e$ - plasma electron temperature at the plasma-sheath interface, $j$ - the total current density in the sheath.

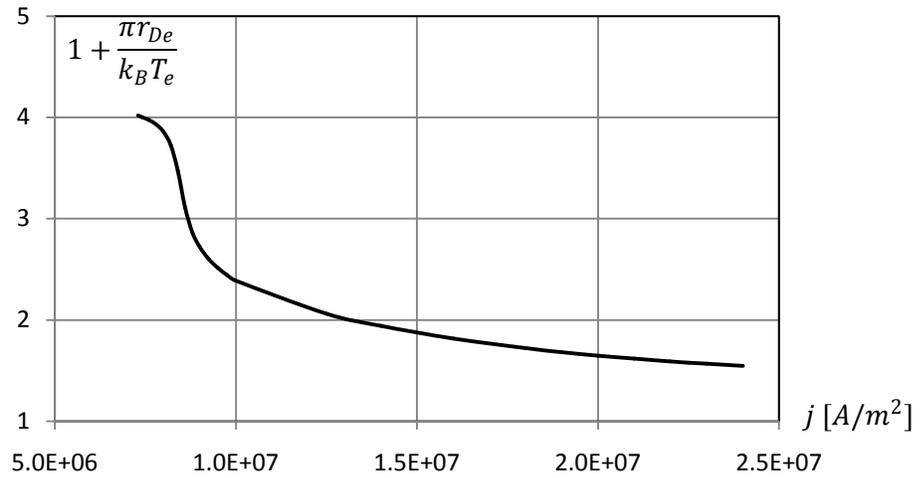

Fig. 11. Model of the cathode spot: Collisional factor vs. total current density in the sheath.